\documentclass[12pt,preprint]{aastex}

\slugcomment{Version \today}
\def\CYAC {$\mathrm{HC_3N}$}     

\def\kms {$\mathrm{km\,s^{-1}}$} 
\def\HII {H{\sc ii}} 
\def\HCFN {$\mathrm{HC_5N}$} 
\def\AMM {$\mathrm{NH_3}$} 
\def\percc {$\mathrm{cm^{-3}}$} 
\def\cmsq  {$\hbox{{\rm cm}}^{-2}$}    
\def\solum {$\hbox{L}_\odot$}

\begin{document}
%
%
   \title{Physical conditions in the Protoplanetary Nebula CRL 618
     derived from observations of vibrationally excited HC$_3$N}

   \author{F. Wyrowski   \altaffilmark{1,2}, 
           P. Schilke        \altaffilmark{2},
           S. Thorwirth \altaffilmark{3}, 
           K.M. Menten \altaffilmark{2}, 
           G. Winnewisser \altaffilmark{3} }

\altaffiltext{1}{Department of Astronomy, University of Maryland,
                 College Park, MD 20742-2421}
\altaffiltext{2}{Max Planck Institut f\"{u}r Radioastronomie,
                 Auf dem H\"{u}gel 69, 
                 D-53121 Bonn, Germany}
\altaffiltext{3}{I. Physikalisches Institut 
                 der Universit\"{a}t zu K\"{o}ln, 
                 Z\"{u}lpicherstra\ss{}e 77, 
                 D-50937 K\"{o}ln, Germany}
%
%
   \begin{abstract} 
     We used the Effelsberg 100~m and IRAM 30~m telescopes to observe
     vibrationally excited cyanoacetylene (\CYAC) in several
     rotational transitions toward the proto-planetary nebula CRL~618.
     Lines from 9 different vibrationally excited states with energies
     ranging up to 1600~K above ground were detected. The lines show P
     Cygni profiles indicating that the \CYAC\ emission originates
     from an expanding and accelerating molecular envelope.  The
     \CYAC\ rotational temperature varies with velocity, peaks at
     520~K, 3~\kms\ blue-shifted from the systemic velocity and
     decreases with higher blueshift of the gas. The column density of
     the absorbing \CYAC\ is 3--6$\times 10^{17}$~\cmsq.  We modeled
     spectra based on spherical models of the expanding envelope which
     provide an excellent fit to the observations, and discuss the
     implications of the models.  Additionally, lines from $^{13}$C
     substituted cyanoacetylene were observed. They can be used to
     constrain the $^{12}$C/$^{13}$C ratio in this source to $10\pm2$.
   \end{abstract}
%

 \section{Introduction}
 At the end of their lives, stars go through a sequence of red giant
 phases accompanied by intense stellar mass loss. This results in
 atomic and molecular circumstellar envelopes enshrouding the central
 asymptotic giant branch (AGB) stars and leading to bright infrared
 sources. Eventually the star evolves into a white dwarf hot enough to
 ionize the envelope, starting the formation of a planetary nebula
 (PN). The transition phase between the AGB and PN stages is very
 short; consequently objects evolving through it, so-called
 proto-planetary nebulae (PPNe), are very rare.  Their physical and
 chemical properties define the starting conditions for planetary
 nebula evolution.  Therefore, it is of importance to determine these
 characteristics observationally.  Reviews on this subject are given
 in Kwok (1993) and Hrivnak (1997).
 
 CRL~618 is the archetypical PPN. Its circumstellar, expanding
 envelope emits in a multitude of molecular lines (e.\ g.\ Bujarrabal
 et al.\ 1988) and contains molecules as complex as benzene, which was
 recently discovered with the Infrared Space Observatory by Cernicharo
 et al.\ (2001); it shows a bipolar morphology/outflow in the optical,
 CO, and HCN emission (Meixner et al.\ 1998; Neri et al.\ 1992); the
 bipolar outflow lobes were imaged at high angular resolution with the
 Hubble Space Telescope (Trammell 2000); a compact, possibly variable
 \HII\ region is just emerging around the central star (Kwok \&
 Bignell 1984; Martin-Pintado et al.\ 1993).
 
To study the physical and chemical properties of the circumstellar
envelope of CRL~618, we started a project to observe the carbon chain
molecules HCN, \CYAC, and \HCFN\ in their vibrationally ground and
excited states. Carbon chain molecules were chosen since they might
be related to the formation of carbon dust grains.
Another objective was that they provide many rotational lines of
different vibrational excitation in a relatively small spectral range,
allowing observations of various lines with equal spatial resolution.
The vibrationally excited levels are populated by IR radiation, and
thus provide the interesting information about the IR fields and
excitation of the gas (Wyrowski et al.\ 1999) which would not be
accessible in the infrared itself due to the high dust extinction.
 
 In this paper, we present an observational study of vibrationally
 excited \CYAC\ in CRL~618 carried out using the Effelsberg 100~m and
 the IRAM 30~m telescopes. In the next section we briefly describe the
 observing techniques. In Sect.~\ref{sec-results}, we summarize our
 main results and in Sect.~\ref{sec-analysis} the data will be
 interpreted in terms of an expanding, molecular circumstellar
 envelope. A discussion of the \CYAC\ analysis comparing our new
 results with complementary molecular studies from the literature is
 then given in Sect.~\ref{sec-discussion}.

  \section{Observations}
   \subsection{Effelsberg 100m}
   We observed CRL~618 with the Effelsberg 100m telescope of the
   Max-Planck-Institut f\"ur Radioastronomie on 1998 April 23.  The
   front end was the facility 7~mm HEMT receiver tuned to the frequency
   of the \CYAC\ $v=0$, $J=5-4$ transition at 45490.313~MHz.  Our
   spectrometer was a 8192~channel autocorrelator which we used with 8
   overlapping subunits of 80~MHz bandwidth to cover a total of
   520~MHz. The resulting spectral resolution was 0.3~MHz after
   smoothing the data to increase the signal-to-noise ratio. The beam
   at the frequencies of the \CYAC\ lines was 20\arcsec. Pointing was
   checked at roughly hourly intervals by means of continuum scans on
   CRL~618 itself. We found the pointing to be accurate to within
   4\arcsec.  Our absolute calibration was based upon continuum scans
   through W3(OH), 3C84, and 3C273, assuming flux densities of 3.5, 6,
   and 25~Jy, respectively, and should be accurate to 25~\%.

   \subsection{IRAM 30m}
   
   The observations with the IRAM 30m telescope were conducted on 1998
   July 17 and 18.  Using the facility SIS receivers simultaneously at
   109.3, 136.6, and 209.4~GHz with half power beam widths of
   22\arcsec, 17.6\arcsec, and 11.5\arcsec, and system temperatures of
   200, 350, and 500~K, respectively. Our spectrometers were an
   autocorrelator with 1633 channels and 0.31 MHz resolution and two
   filter-banks with 1~MHz resolution and 512 channels. All lines were
   observed in the lower sideband with typical sideband rejections of
   0.003, 0.025, and 0.04, at 109.3, 136.6, and 209.4~GHz,
   respectively. The wobbling secondary mirror was used with a beam
   throw of 200\arcsec\ and a frequency of 0.25~MHz, resulting in
   spectra with flat baselines. The antenna main beam efficiencies,
   measured on Mars, were 0.63, 0.51 and 0.45, at 109.3, 136.6, and
   209.4~GHz, respectively. We estimate that our absolute calibration
   uncertainy is lower than 25~\%.

  \section{Results}
  \label{sec-results}
   \subsection{The 100m observations near 45~GHz}
\begin{figure}[p]
      \epsscale{.7}
      \plotone{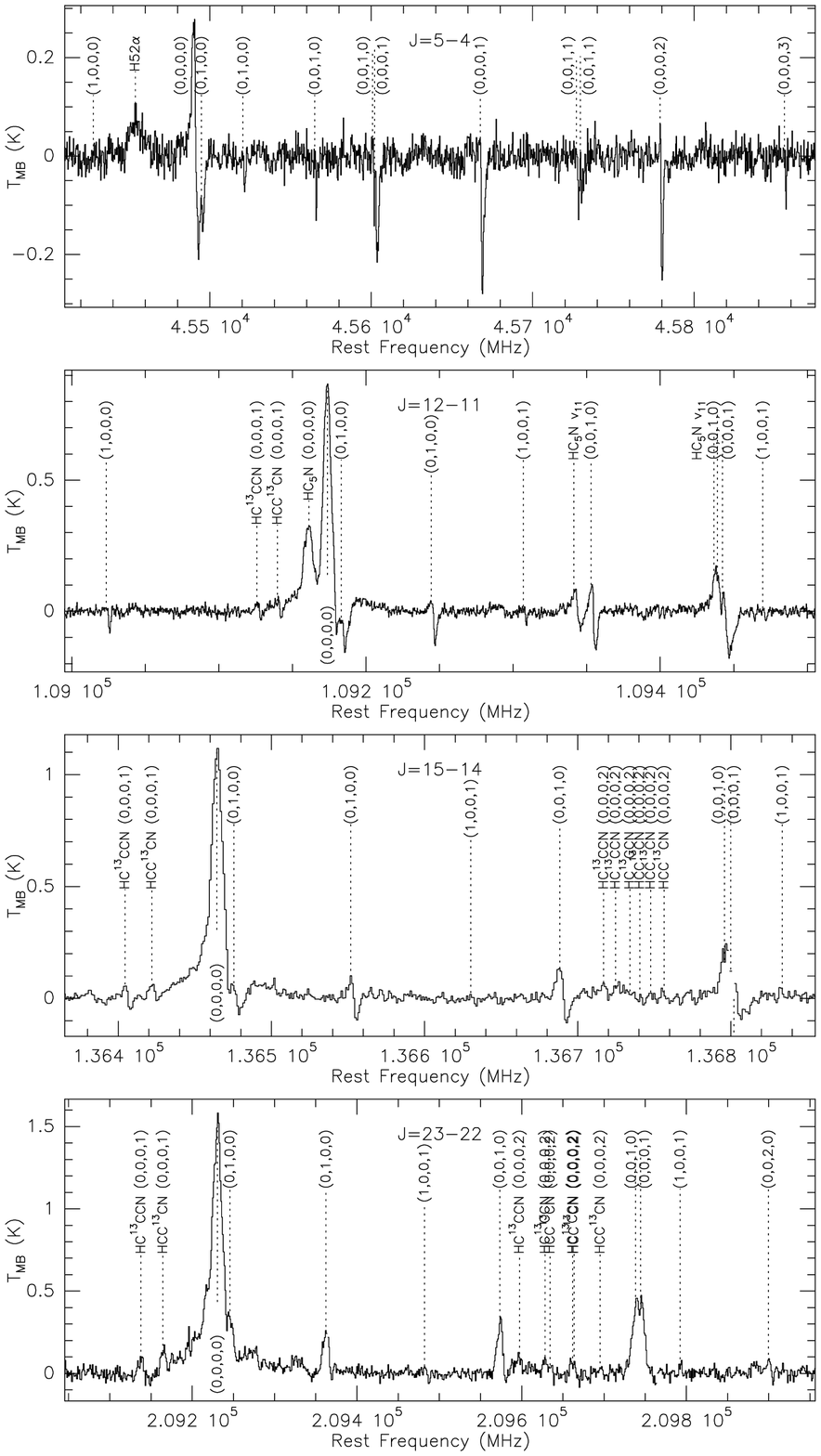}
    \caption{
      \CYAC\ spectra toward CRL~618.
      }
    \label{all-hc3n}
   \end{figure}

\begin{figure}[p]
      \epsscale{.8}
      \plotone{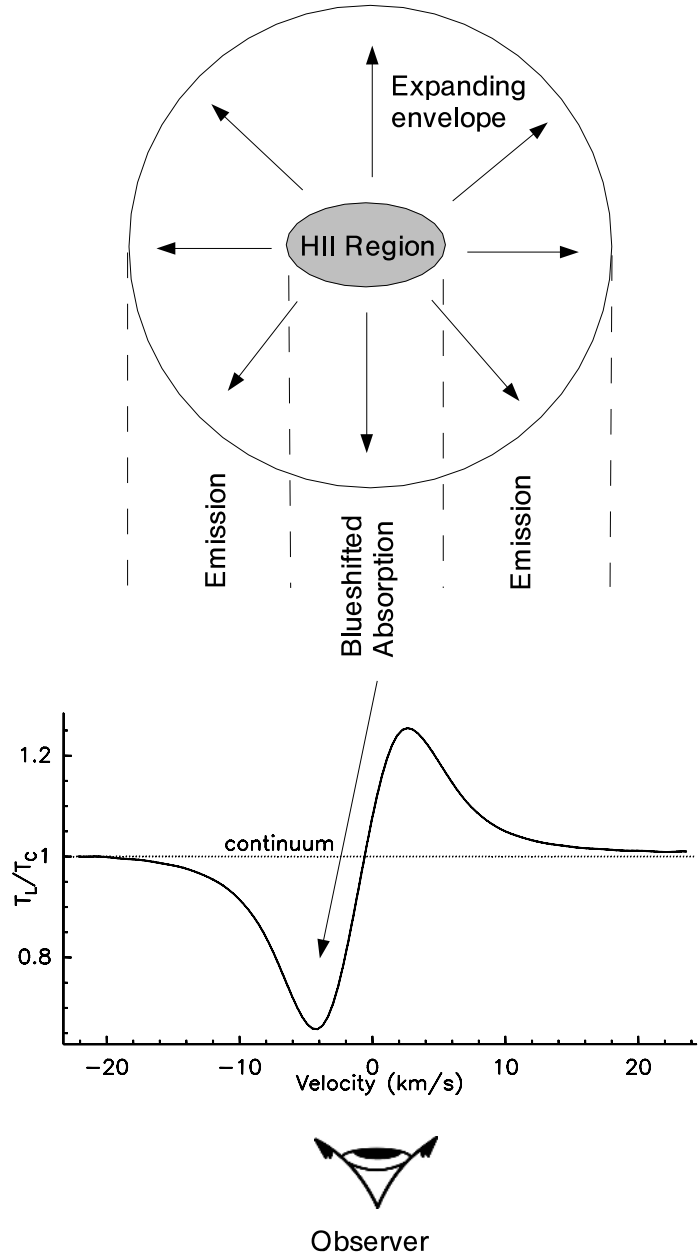}
    \caption{
      General sketch illustrating the P Cygni line profile formation in an
      expanding circumstellar envelope. The absorption feature is due
      to the blueshifted part of the envelope in front of the strong
      continuum from the \HII\ region.
      }
    \label{sketch}
   \end{figure}

   The integrated, baseline subtracted \CYAC\ $J=5-4$ spectrum of
   CRL~618, merging the different correlator subunits into one
   spectrum, is shown in the upper panel of Fig.~\ref{all-hc3n}.  The
   vibrational ground state line of cyanoacetylene shows a P Cygni
   profile indicating that the line originates from a hot, expanding
   circumstellar envelope close to the exciting star. This situation
   is illustrated in Fig.~\ref{sketch}.
   In all the
   vibrationally excited lines the emission part is too weak to be
   detected and only absorption is seen. We have detected pure
   rotational transitions within all fundamental bending modes
   $\nu_7$, $\nu_6$, $\nu_5$, the $\nu_6\nu_7$ combination mode and
   the $2\nu_7$ and $3\nu_7$ overtones. The line parameters from
   Gaussian fits to the lines are given in Table~\ref{param-100m}.
   The emission and absorption features of lines showing P Cygni
   profiles were fitted separately. The brightness of emission lines
   is given in main beam brightness temperature $T_{\rm MB}$. We
   report the strength of absorption features in $T_{\rm MB}/T_{\rm
   C}$, with $T_{\rm C}$ as the continuum temperature, which is
   proportional to the optical depth of the lines for $T_{\rm
   MB}/T_{\rm C}$ much smaller than unity.  Another advantage of the
   $T_{\rm MB}/T_{\rm C}$ scale is its independence on the telescope
   beam size, since the size of the continuum source is considerably
   smaller then the beam (0.1$\times$ 0.4 arcsec, Martin-Pintado et
   al.\ 1993), as for the millimeter observations described in the
   next section.

   \subsection{The IRAM 30m observations}
   The lower three panels of Fig.~\ref{all-hc3n} show the spectra of
   \CYAC\ $J=12-11$, $J=15-14$, and $J=23-22$.  At the frequencies of
   $J=12-11$ and $J=15-14$, even more lines lines show P Cygni
   profiles. Comparison of, for example, all the $v_5=1f$ lines
   demonstrates the continuous transition from absorption to emission
   of the lines at higher frequencies.  Vibrational satellites
   corresponding to six different modes of vibration of \CYAC\ can be
   identified: Besides the ground vibrational state, all fundamental
   bending modes $\nu_7$, $\nu_6$, $\nu_5$ are detected as in the case
   of $J=5-4$.  Additionally the stretching vibration $\nu_4$ and the
   combination mode $\nu_7 \nu_4$ were detected. At 1.3mm ($J=23-22$),
   also the $2\nu_6$ overtone is detected.  The latter three
   vibrational modes correspond to energies from 1300 to 1600~K above
   ground (see vibrational energy levels in Wyrowski et al.\ 1999).
   All the line parameters are given in Tables~\ref{param-30m-12},
   \ref{param-30m-15} and \ref{param-30m-23}.  Only heavily blended
   lines were omitted.  Several rotational lines of $^{13}$C
   isotopomers of \CYAC\ in their bending modes $\nu_7$ and $2\nu_7$
   are observed as well. For all the identifications and analyses we
   used the improved laboratory frequencies of Thorwirth et al.\
   (2000, 2001).  Moreover, the 109~GHz spectrum shows the $J=41-40$
   ground state line of \HCFN\ along with its $v_{11}=1e$ vibrational
   satellite.  A study of this molecule in CRL~618 is discussed in
   Thorwirth (2001).

   \subsection{The continuum emission of CRL~618}
   \label{sec-continuum}
   
\begin{figure}[p]
      \epsscale{1}
      \plotone{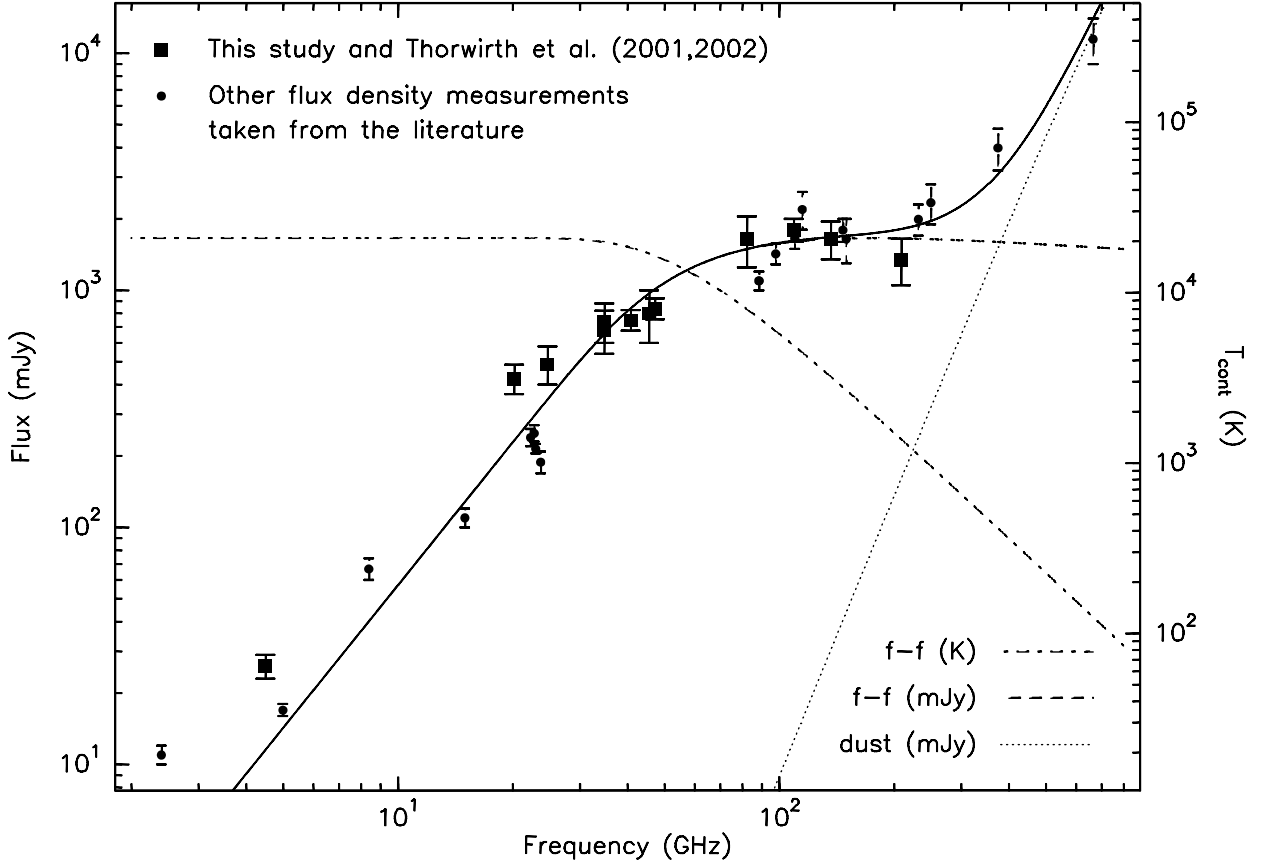}
    \caption{
      Continuum spectrum of CRL~618. The fit of free-free emission to
      the flux values obtained in this study is shown as a dashed
      lines. Dust emission is shown as a dotted line and the sum of
      both components as a solid line. Literature flux densities were
      taken from Kwok \& Bignell (1984), Turner \& Terzian (1984),
      Martin-Pintado et al.\ (1988), Neri et al. (1992), Knapp,
      Sandell, \& Robson (1993), Martin-Pintado et al.\ (1993),
      Yamamura et al.\ (1994), Hajian, Phillips, \& Terzian (1995),
      Martin-Pintado et al.\ (1995), Knapp et al.\ (1995), Meixner et
      al.\ (1998). Additionally, the free-free emission is given in
      a temperature scale (see Sect.~\ref{sec-temp}).
      }
    \label{compare-cont}
   \end{figure}

   With the MPIfR 100m telescope, 
   we observed drift scans centered on CRL~618 to
   obtain its continuum flux density. With the 30m, we estimated
   millimeter-wavelength continuum flux densities of CRL~618 from the
   total power offsets of the measured spectra, leading to consistent
   results between different scans. This method is reliable since we
   used the chopping secondary mirror so that instrumental and
   atmospheric drifts are kept small.  The measured fluxes densities
   at all observed frequencies are given in
   Table~\ref{cont-fluxes}. In addition, we use the flux densities
   obtained during our studies of $\rm HC_5N$ and HCN in CRL~618
   (Thorwirth 2001, Thorwirth et al.\ 2002). In
   Fig.~\ref{compare-cont} we compare the measured fluxes with
   continuum fluxes from the literature.  At millimeter wavelengths
   flux density estimates at similar frequencies agree with eachother
   within a factor of two, but for longer wavelengths the spread is
   larger.  The observations below 50~GHz were done using the VLA, the
   100m and the Arecibo telescope. There is no trend for
   interferometrically determined flux densities being smaller, so
   that we don't think the interferometer observations miss flux due
   to the lack of short spacings. There is also no clear trend for
   long term variability: the flux measurements by Turner \& Terzian
   (1984) at 2.4~GHz and by Knapp et al.\ (1995) at 8.4~GHz were
   performed in 1981 and 1991, respectively, and are in accordance
   with the recent VLA 5~GHz value by Thorwirth et al.\ (2002). On the
   other hand, variability on shorter timescales might be the case:
   Kwok \& Feldman (1981) reported a brightening of CRL~618 by a
   factor of 2 over a 2-3 yr interval at frequencies around
   10~GHz. For the following 5 years then, no significant increase
   could be detected (Martin-Pintado et al.\ 1988). Here we only model
   the continuum emission for the frequency range of 45--210~GHz
   covered in the present study, assuming constant flux densities in
   the short time interval from April to July 1998.  The fit to the
   continuum flux densities in Fig.~\ref{compare-cont} was obtained by
   calculating the emerging flux from a homogeneous, circular \HII\
   region. The radius was fixed to 0.11\arcsec\ (or 190 AU at a
   distance of 1.7~kpc, Westbrook et al.\ 1975) to agree with the area
   of the \HII\ region obtained by Martin-Pintado et al.\ (1995). Then
   the electron temperature $T_e$ and emission measure $EM$ were
   chosen to fit the data.  Our estimate for $EM$ agrees with the
   result by Martin-Pintado et al.\ (1988). $T_e$ from our fit is 50\%
   higher.

  \section{Analysis}
  \label{sec-analysis}
  \subsection{Velocities}
  
\begin{figure}[p]
      \epsscale{1}
      \plotone{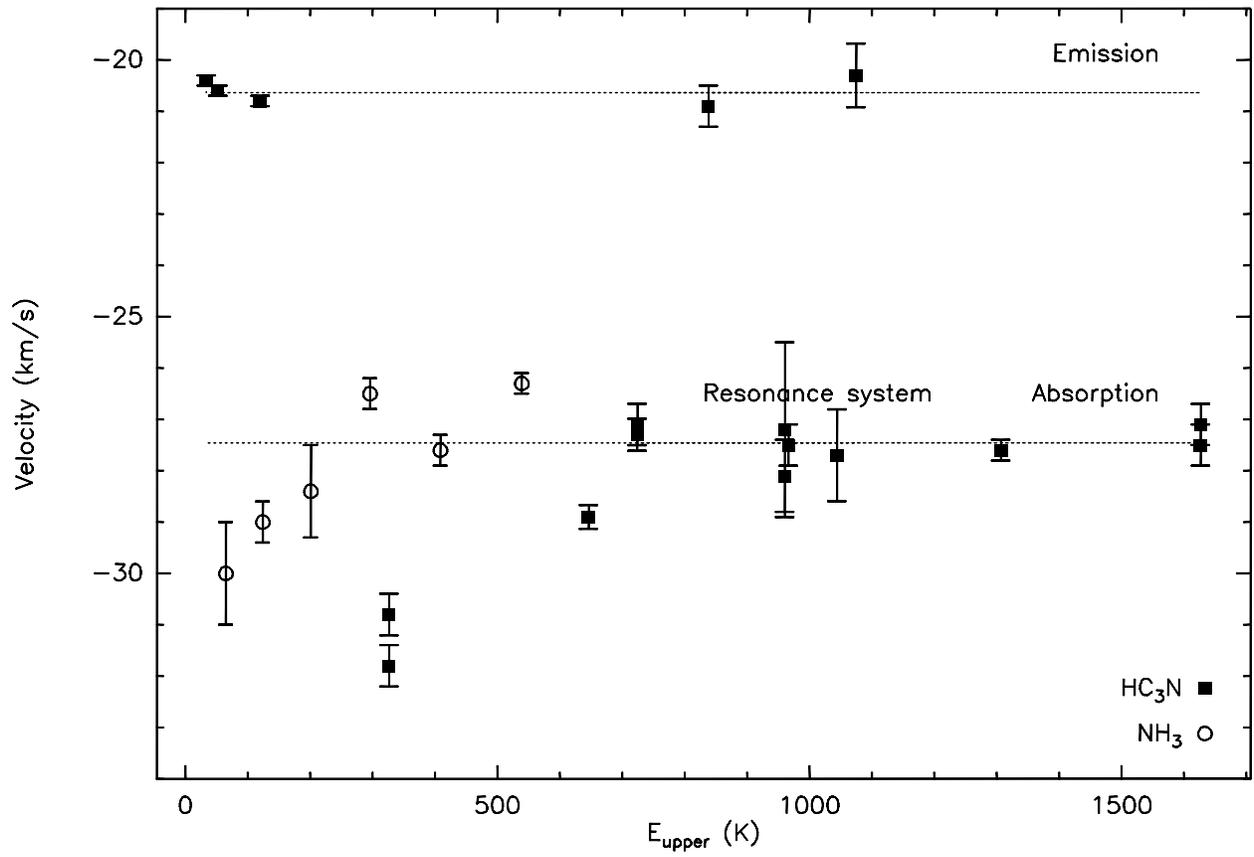}
    \caption{
      Velocities from the Gaussian fits to the \CYAC\ lines as a
      function of the upper energies of the transitions.
      Data from the $v_5=1/v_7=3$ Fermi resonance system are at about
      1000~K.
      }
    \label{show_vel-all}
   \end{figure}

  In Fig.~\ref{all-hc3n}, the frequencies of all \CYAC\ lines are
  marked assuming a velocity of the gas of --20.5~\kms. For the emission
  lines this is a good assumption but it can be seen easily that all
  the absorption features are blue-shifted. This situation is shown in
  more detail in Fig.~\ref{show_vel-all} for all unblended lines with
  either mostly emission or absorption: for the emission lines, there
  is little variation of velocity with upper energies of the
  transitions and their average velocity is $-20.6\pm 0.2$~\kms. For
  the absorption lines the situation is more complicated.  Line
  velocities below 1000~K show a rising blueshift with decreasing
  energies above ground. Above 1000~K, the velocities of the
  absorption is almost constant with a velocity of $-27.2\pm
  0.2$~\kms.
  In the framework
  of an expanding cooling envelope this can be interpreted as lower
  energy lines absorbing mainly from cooler parts of the envelope and
  higher energy lines from hotter parts closer to the exciting star. A
  constant velocity for lines higher than 1000~K means, that no \CYAC\
  is found closer to the star, hence marking the radius of the
  envelope where the temperature gets small enough that \CYAC\ is more
  effectively produced than destroyed. 

  For comparison, \AMM\ line velocities taken from Martin-Pintado \&
  Bachiller (1992) are shown in Fig.~\ref{show_vel-all} as well. A
  rising blueshift with decreasing energies can also be seen in \AMM\ 
  but shifted to lower energies above ground. Addtionally,
  Martin-Pintado et al.\ (1992) found another absorption component at
  about $-$56~\kms, which was later interpreted (Martin-Pintado et al.\ 
  1993, 1995) as being due to postshocked gas from the interaction of
  the high velocity outflow with the expanding AGB envelope. In this
  picture ammonia absorption is found symmetrically at $\pm$ 15~\kms\ 
  from the terminal velocity of the unshocked AGB envelope at
  --40~\kms. This situation is illustrated in Fig.~5 of Martin-Pintado et al.\ 
  (1995).  In contrast, the \CYAC\ absorption profiles are single
  peaked, suggesting that \CYAC\ mainly resides in the unshocked gas.
  It follows then that the ammonia HC component (Martin-Pintado et
  al.\ 1992) shown in Fig.~\ref{show_vel-all} is a blending of the
  redshifted lobe of the postshock material with the hot unshocked
  inner part of the AGB envelope, seen in \CYAC.

  \subsection{The high-velocity outflow}
  
  The lines of the ground vibrational state of \CYAC\ show broad lines
  wings. Hence some \CYAC\ is found in the 200~\kms\ molecular outflow
  (Cernicharo et al.\ 1989; Neri et al.\ 1992). We fitted the line
  wings by masking all blending lines and the fit results are given in
  Tables 1--4. The lines with higher excitation do not show any
  detectable wings, maybe with the exception of the $J=5-4, v_7=1f$
  line. Assuming optically thin emission from the \CYAC\ line wings,
  the rotational temperature of the high velocity gas is $60\pm 7$~K,
  where we included the \CYAC\ results of Cernicharo et al.\ (1989)
  into the fit. The corresponding \CYAC\ column density averaged over
  a 22\arcsec\ beam is $46\pm 9 \times 10^{12}$~\cmsq.

  \subsection{Temperatures}
   \label{sec-temp}
  
   A rough estimate of the cyanoacetylene level populations can be
   obtained by using the lines with dominant emission features and
   plotting the column densities in the levels against energy above
   ground in a Boltzmann diagram assuming optically thin emission.
   This leads to a temperature of $280\pm 20$~K which should be
   considered as the average over all distances from the central star.
   Since the optical depth of the emission lines is not known a
   priori, this temperature is only an upper limit.
  
  From Fig.~\ref{show_vel-all}, it is clear that absorption lines with
  velocities of --28 to --27~\kms\ trace the hottest component of the
  gas on the line of sight in front of the continuum emission. Using
  only these vibrationally excited lines, we estimate a temperature of
  $520\pm 80$~K from a Boltzmann plot. Since the measured quantity
  $T_{\rm MB}/T_{\rm C}$ is approximately $1-\exp(\tau)$ (for $T_{\rm
  C}>>T_{\rm ex}$), the optical depth is known and cannot simulate a
  higher temperature.  Fig.~\ref{compare-cont} shows $T_{\rm C}$ as a
  function of frequency and it can be seen that for all frequencies
  considered in this study $T_{\rm C}>1000$~K. The reason for the
  decrease in absoption with increasing frequency is that the emission
  gets stronger and fills up the absorption part of the line profile.
  
\begin{figure}[p]
      \epsscale{1}
      \plotone{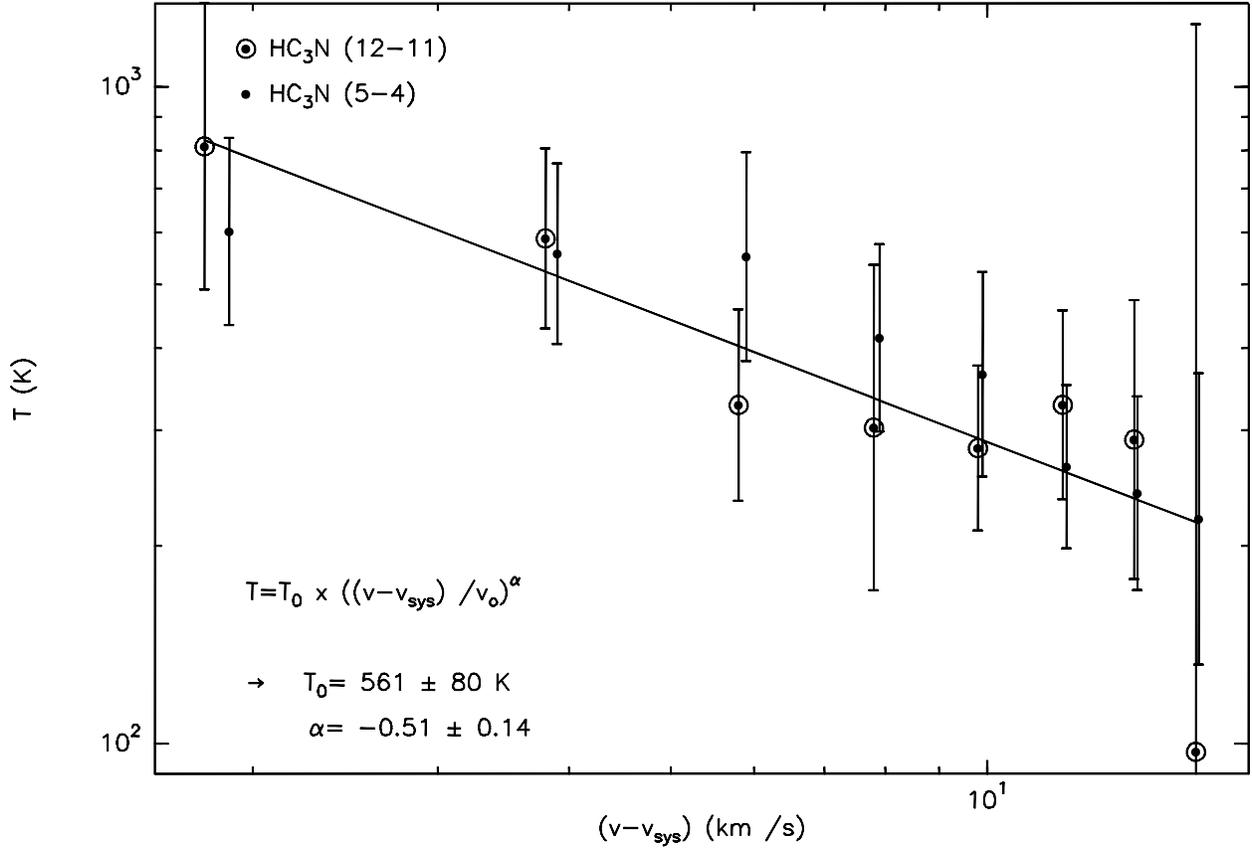}
    \caption{
      Temperatures estimates from Boltzmann plots of absorption
      detected in different velocity bins. The temperatures have been
      determined for \CYAC\ (5--4) and (12--11) separately to test the
      reliability of the method. The solid shows a fit for the power
      law temperature distribution discussed in Sect.~\ref{sec-model}.
      }
   \label{rd-velint}
  \end{figure}

  In a refinement of this method, we divided all the absorption lines
  into velocity bins of 2~\kms\ and estimated the temperature in each
  bin from Boltzmann plots separately. The result is shown in
  Fig.~\ref{rd-velint} for \CYAC\ (5--4) and (12--11), using a
  systemic (stellar) velocity $v_{\rm sys}$ of --24.2~\kms\ (see
  discussion in the next section). A clear increase of the temperature
  with lower velocities, presumably closer to the central star, is
  evident from the population of the vibrationally excited states in
  both rotational levels, confirming again the interpretation that the
  lines originate from an expanding cooling envelope.

  \subsection{A spherical LTE model for the envelope}
   \label{sec-model}

  To account for the observed P Cygni profiles, a more detailed
  approach is necessary. We therefore modeled an expanding
  spherical envelope with radial power laws for temperature,
  density, and velocity:
  \begin{equation}
   T=T_0\,(r/r_0)^{\alpha_T}, \quad 
   n=n_0\,(r/r_0)^{\alpha_n}, \quad 
   v=v_0\,(r/r_0)^{\alpha_v}
   \end{equation}
   For the \HII\ region in the center we used the parameters derived
   from our continuum observations in Sect.~\ref{sec-continuum}. We
   then calculated for a given molecular line and velocity the
   resulting intensity distribution assuming LTE conditions (see
   Appendix~\ref{sec-rt}), and from that the emerging spectra in our
   observing beams. Additional free parameters are the systemic (stellar)
   velocity $v_{sys}$ of the envelope,  the intrinsic line width
   $\Delta v$, and the $\rm ^{12}C/^{13}C$ ratio.  For the inner
   radius $r_0$ we used the radius of the \HII\ region. 

   Our search for the best fit parameters was guided by results of the
   previous subsections: the velocity at the inner radius $v_0+v_{\rm sys}$
   was chosen to lie close to the observed velocity of the high energy
   absorption lines (Fig.~\ref{show_vel-all}) and the temperature at
   the inner radius has to be of the order of the temperature found
   from the Boltzmann plot of these lines. As a more quantitative
   approach, we fit the temperature distribution shown in
   Fig.~\ref{rd-velint} with the power law given above and adjusted
   $v_{\rm sys}$ by minimizing the difference between fit and data, leading to
   $v_{\rm sys}=-24.2$~\kms.  The resulting $T_0$ is close to the value
   estimated from the high energy absorption lines. The resulting
   power law index $\beta$ of --0.5 constrains the ratio
   $\alpha_T/\alpha_v$. The ratio $\alpha_n/\alpha_v$ is best
   determined by the total widths of the P Cygni profiles, since it
   mainly governs out to which velocities absorption/emission is still
   detected.
   Finally, the ratio
   between emission and absorption peaks depends on $\alpha_v$,
   which determines the actual spatial size of the emission, hence a
   large $\alpha_v$ leads to small sizes and hence less emission in
   the modeled spectra. 
  
   At this point, it is necessary to discuss an important limitation
   on the accuracy of the \CYAC\ column density determination: the
   exact temperature dependence of the \CYAC\ partition function, including
   both the rotational and the vibrational part, is unknown. In a true
   LTE approximation, all existing energy states have to be included,
   which leads for \CYAC\ due to its relatively low vibrationally
   excited states to a sharp increase of the partition function for
   temperatures above about 300~K. Collisional rates for the
   excitation of the vibrationally excited states are mostly unknown
   and the measurements and estimates that exist suggest a fast
   increase of the resulting critical densities with energies above
   ground (Wyrowski et al.\ 1999). On the other hand, most of the
   lines might be populated rather by infrared pumping than
   collisions. To reflect these uncertainties, we determined the
   column density using the two possible extreme cases: a) counting
   all the states in the partition function and b) counting only the states
   up to the highest observed energy above ground.
   
   For the fit we used only unblended lines with small errors on the
   line frequencies.  The resulting best fit parameters are given in
   Table~\ref{fit-results}: The temperature exponent and the inner
   temperature are close to the values found by Meixner et al.\ (1998)
   for CIT~6 and CRL~618. It is clearly hotter and steeper than the
   temperature structure of an optically thin cloud heated by a
   central source ($T_0$=420~K, $\alpha_T$=--0.4).  This is not
   unexpected for the dense, dusty inner part of the CRL~618 envelope:
   if the dust is still optically thick in the NIR, the envelope heats
   itself to higher temperatures by absorbing its own thermal
   emission, leading to higher temperatures in the inside and a
   steeper temperature structure (cf.\ Sect.~\ref{sec-dusty}). This
   power law can also be compared with results from modeling the
   envelope of IRC+10216, where Mamon et al.\ (1998) found an inner
   power law index of --0.72 by fitting the model results of Kwan \&
   Linke (1982). Millar et al.\ (2000) give a temperature index of
   --0.79 for IRC+10216 in excellent agreement with our results toward
   CRL~618.

   For a
   spherical wind, the equation of continuity requires a density
   exponent of $-(2+\alpha_v)$. We model a steeper density exponent
   than this for both forms of the partition function. Since the
   \CYAC\ density is simply $n_{\rm H_2}$ times $X$(\CYAC), the \CYAC\
   abundance, this probably means that in the flow \CYAC\ is destroyed
   or converted into more complex molecules.
   
\begin{figure}[p]
      \epsscale{1}
      \plotone{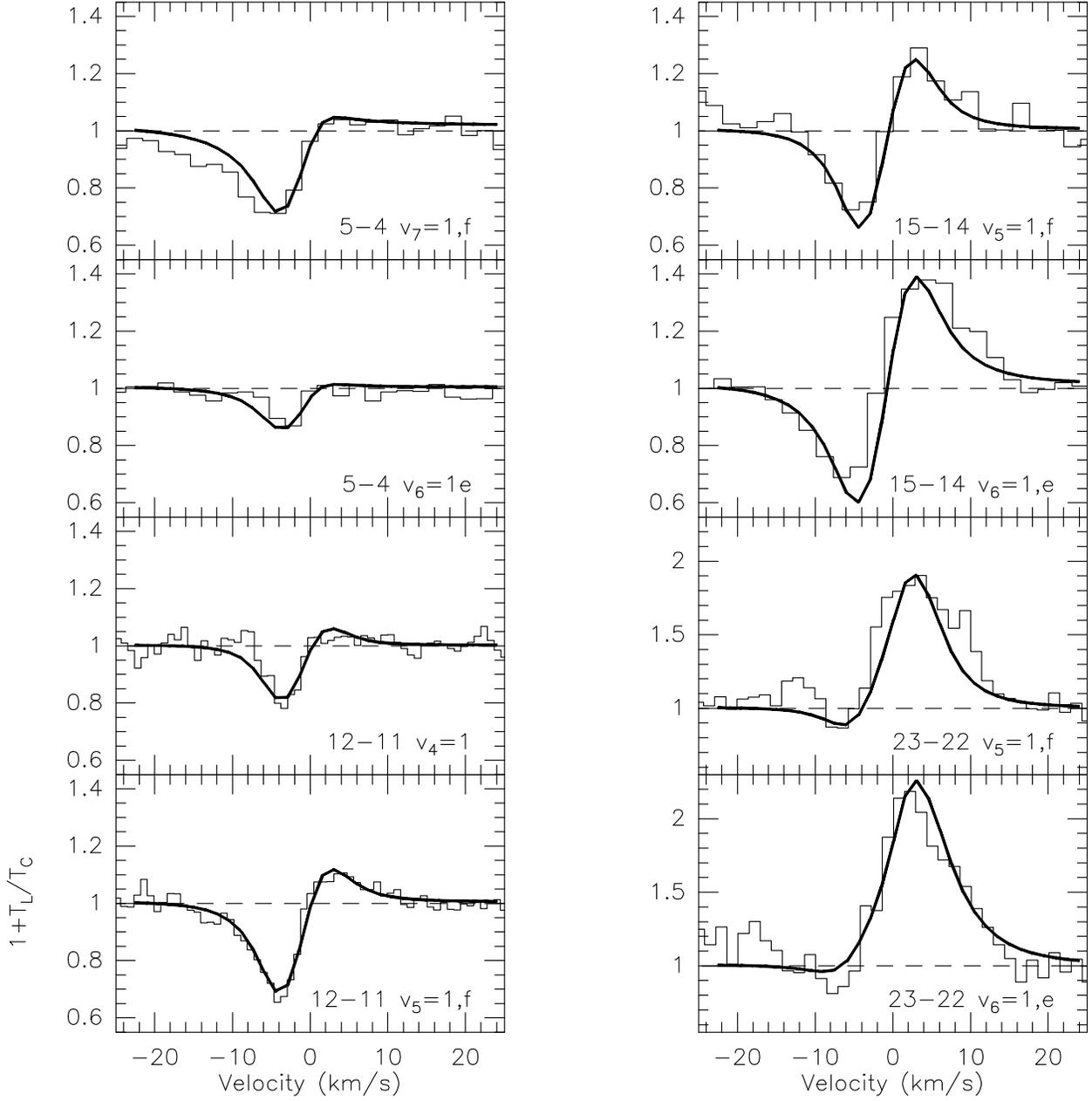}
    \caption{
      Comparison between observations (histogram) and line profiles
      from the model described in Sect.~\ref{sec-model}. The velocity
      scale is relative to the systemic velocity.
      }
   \label{3d-models}
  \end{figure}

   Fits to data are shown in Fig.~\ref{3d-models}.  Most lines show
   excellent agreement between data and model, which is very
   surprising given the simple assumptions used.  The remaining small
   differences can probably be attributed to violations of the
   assumptions, notably local thermodynamical equilibrium and spherical 
   symmetry (CRL~618 does have a bipolar flow).

  \subsection{Exact modeling of the dusty circumstellar envelope}
   \label{sec-dusty}

\begin{figure}[p]
      \epsscale{1}
      \plotone{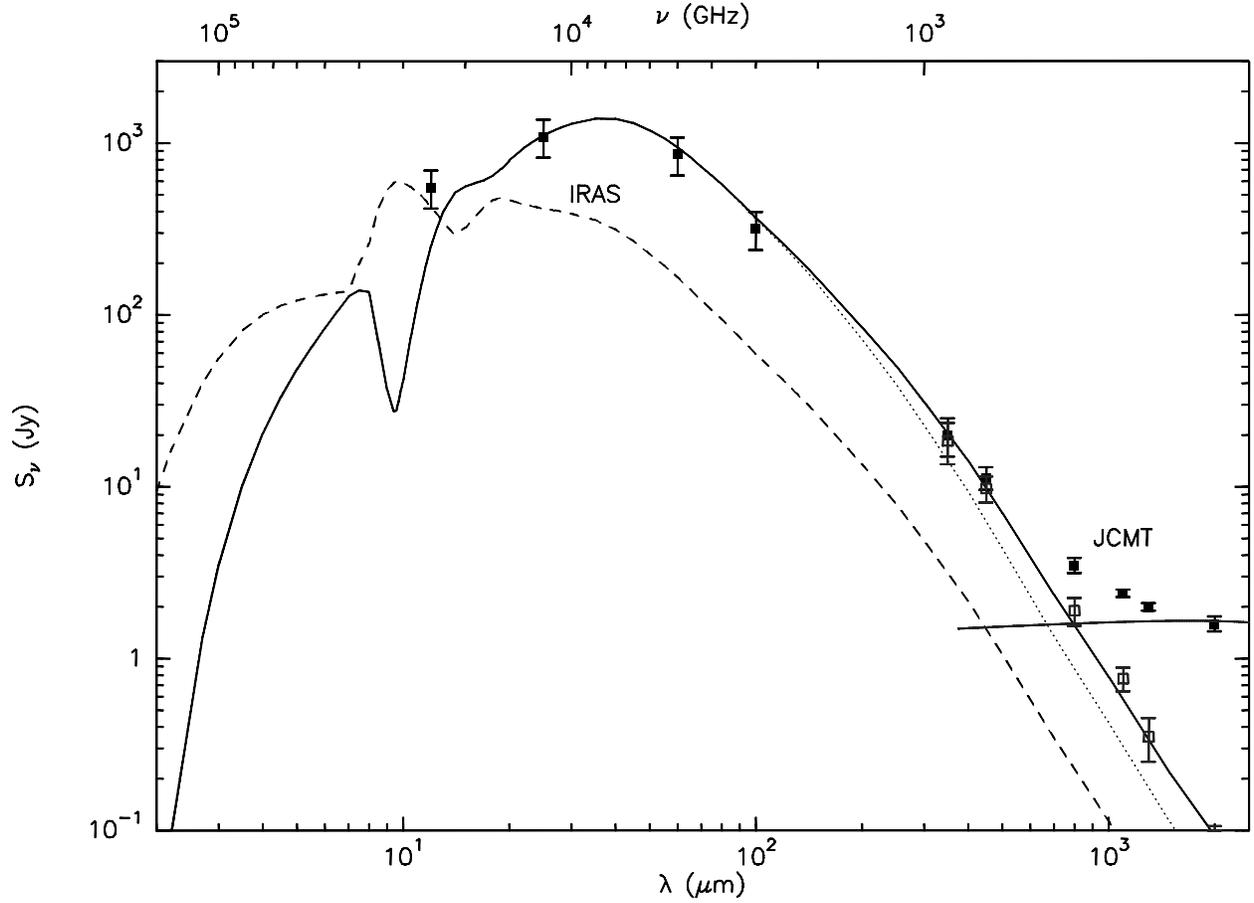}
    \caption{
      DUSTY models of the CRL~618 SED compared with IRAS and SCUBA
      data (Knapp, Sandell \& Robson 1993, unfilled squares are
      corrected for free-free flux shown with a separate solid
      line). The solid line shows a model with $\tau_{\rm V}=100$ and
      thickness s$=3000$. The dotted line is the same model with
      s$=1000$ and the dashed line a model with $\tau_{\rm V}=10$.
      }
   \label{dusty-sed}
  \end{figure}

   The results of our LTE analysis of the envelope in the last section
   can be compared with exact modeling of the envelope using
   calculations of the radiative transport in dusty shells (DIRT:
   http://dustem.astro.umd.edu, Wolfire et al.\ 1986; DUSTY, Ivezi{\'
   c} et al.\ 1999). For the DUSTY modeling we used a 30000 K black
   body as radiation source, a mixture of Draine \& Lee (1984)
   silicates and graphite with a standard MRN grain size distribution,
   and a dust temperature of 800~K at the inner boundary. The density
   structure was derived within DUSTY from full dynamics
   calculations. The remaining free parameters are the optical depth
   and the thickness of the envelope. Fig.~\ref{dusty-sed} shows the
   model results compared to measured IRAS and JCMT flux densities
   (Knapp, Sandell \& Robson 1993).  The optical depth $\tau_{\rm V}$
   determines the overall shape of the SED, whereas the thickness
   s=r(out)/r(in) only modifies the submm part of the SED.  We
   subtracted the free-free contribution originating from the HII
   region (cf.\ Sect.~\ref{sec-continuum}) from the submm
   emission. The SED shape of the DUSTY models is independent of the
   luminosity, which just scales the model fluxes up or down. An
   adequate fit to the data is reached using $\tau_{\rm V}=100$, a
   thickness s$=3000$ and a luminosity of $2\times 10^4$~\solum. r(in)
   is the dust sublimation radius which is for this model at 0.11
   arcsec, hence this is the same radius we used in the last
   section. The outer radius r(out) of 330 arcsec is comparable to the
   400 arcsec observed by Speck et al.\ (2000) with ISO. As seen in
   the Figure, a smaller outer radius can reproduce the submillimeter
   flux densities.

   While these results are interesting in their
   own right, for our interpretation of the vibrationally excited
   \CYAC\ lines only the inner part of the envelope is relevant and
   especially the high optical depths of the envelope. We checked the
   inner temperature structure (thickness s$<$10) of DUSTY and DIRT
   models with the high optical depth to fit the SED and found that
   the temperature falls off considerably faster than in the optical
   thin case. While the exact structure of the models differ 
   they can roughly be
   approximated with $\alpha_T$=--0.8 power laws, hence consistent with
   the results of the last section.

   Since DUSTY allows us also to model the envelope expansion by assuming
   that it is driven by radiation pressure on the dust grains
   (Ivezi{\' c} \& Elitzur 1995, Elitzur \& Ivezi{\' c} 2001), we can
   compare the velocity structure predicted by DUSTY with our results
   from the last section. The model used to fit the SED leads to a
   terminal wind velocity $v_\infty$ of 4~km/s for a gas-to-dust ratio
   $r_{\rm gd}=200$ and a dust grain bulk density
   $\rho_s=3$~g/cm$^3$. $v_\infty$ scales with $(r_{\rm
   gd}\rho_s)^{-1/2}$ (see Elitzur \& Ivezi{\' c} 2001 for details).
   Hence, explaining a $v_\infty$ of order 20~km/s, as implied by our
   data, would require very unlikely values for $r_{\rm gd}$ and/or
   $\rho_s$. But even then, the P Cygni profiles predicted by a DUSTY
   velocity structure would be very broad and less peaked than the
   observed ones. The reason is that for radiatively driven, optically
   thick winds most of the acceleration takes place in a very thin
   part of the shell: 90\% of $v_\infty$ is reached already at about
   2 $r_{\rm in}$. We therefore conclude that the expansion of the
   CRL~618 envelope cannot exclusively be due to radiation pressure
   and other processes have to be invoked, e.\ g.\ processes related
   to the powerful outflow associated with CRL~618.

  \subsection{The total absorbing column density and the $\rm
   ^{12}C/^{13}C$ ratio}

   The observed absorption lines can be used to estimate the \CYAC\
   column density on the line of sight. As discussed in the last section,
   the major uncertainty here is the partition function. We use again
   two extreme cases to constrain the column density: using a temperature
   of 520~K (Sect.~\ref{sec-temp}) the \CYAC\ column density is 3--6$\times
   10^{17}$~\cmsq.

\begin{figure}[p]
      \epsscale{1}
      \plotone{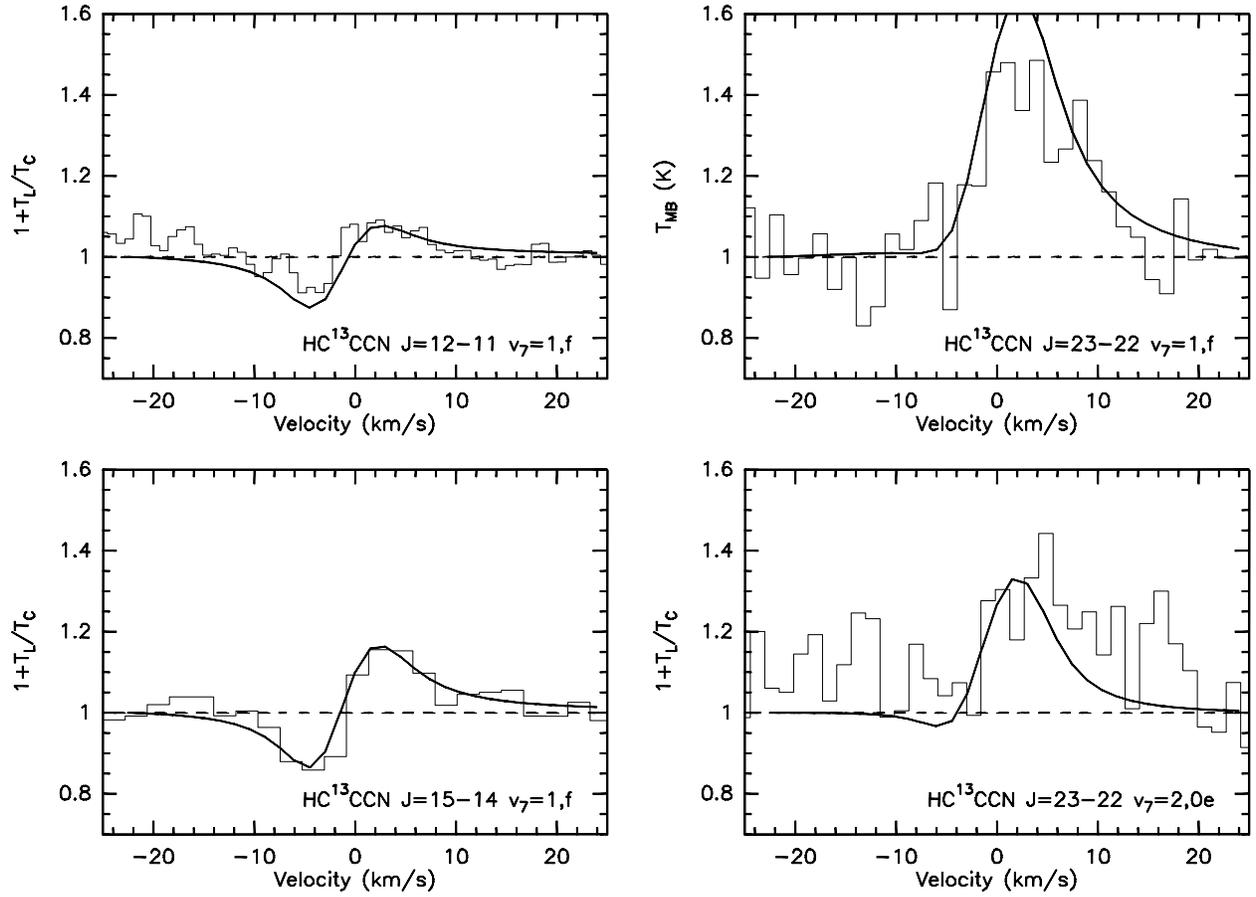}
    \caption{
      Comparison between observations (histogram) and line profiles
      from the model described in Sect.~\ref{sec-model} for lines of
      $\rm HC^{13}CCN$.
      }
   \label{3d-13c}
  \end{figure}

   To estimate the $\rm ^{12}C/^{13}C$ ratio to a high accuracy, we
   determined the minimum $\chi^2$-sum between data and model for the
   \CYAC\ $v_5=1f$ and the $\rm HC^{13}CCN$  $J=15-14$ $v_7=1f$ lines.
   These lines were chosen, because they appear to be unblended and
   were observed simultaneously with the same receiver. Hence, in the
   ratio, all calibration uncertainties cancel out. Appearing mostly in
   absorption, these 3mm lines are much more sensitive to column density
   than the corresponding $\rm ^{12}C,^{13}C$ lines at the higher frequencies.
   The errors in the
   resulting densities were determined by varying the densities around
   the best fit values. This leads to a $\rm ^{12}C/^{13}C$ ratio of
   $10 \pm 2$ and the model fits to all of the $\rm HC^{13}CCN$ lines
   are shown in Fig.~\ref{3d-13c}. 

  \section{Discussion and Conclusions}
  \label{sec-discussion}
  Our results include the best determination of the temperature
  structure in CRL~618 so far. The method we used offers several
  advantages: (i) highly excited absorption lines of \CYAC\ in
  vibrationally excited states are able to probe the hottest, inner
  part of the molecular envelope in a pencil beam. Since vibrationally
  excited lines with the same $J_u$ are observed simultaneously, their
  relative calibration is excellent and the peak temperature found is
  $520\pm 80$~K. (ii) Due to the expansion of the envelope, the
  temperature structure can be determined as a function of velocity.
  (iii) From the emission part of the observed P Cygni spectra the
  spatial extend of the envelope at a certain velocity can be
  determined, which leads then to temperature structure as a function
  of radius ($T\propto r^{-0.8}$).

  How does our result compare with previous measurements? Cernicharo
  et al.\ (2001) find a temperature of 200~K from their ISO
  mid-infrared observations of HCN and $\rm C_2 H_2$.  They were not
  able to resolve the velocity structure and therefore probe with
  their absorption lines an average over the whole line of sight,
  hence a lower temperature. In the analysis of far-infrared ISO
  observations of CRL~618 Herpin \& Cernicharo (2000) distinguish
  between different shells close to the central star, with
  temperatures of 1000, 800 and 250~K, decreasing away from the star
  with radii of 0.3, 0.5 and 0.8 arcsec, respectively. The shells with
  the higher temperatures than our estimate seem to probe a photon
  dominated region, where \CYAC\ might not be able to exist yet.

  Exact modeling of the dusty circumstellar envelope using DUSTY was
  able to reproduce the temperature structure and the spectral energy
  disribution, but failed to predict the observed velocity structure
  of the envelope, indicating that radiation pressure alone is not
  sufficient to drive the expansion.

Isotopic abundance ratios in AGB stars deliver, in principle, valuable
information on nuclear processes and dredge-up convective processes
that bring up enriched material into the stars' outer layers from
where they finally end up in the circumstellar envelopes (see,
e.g. Balser, McMullin \&\ Wilson 2002 and references therein).  The
${\rm HCCCN}/{\rm HC}^{13}\rm{CCN}$ abundance ratio of 10 we determine
is much lower than the solar system value of 89.  Our ratio is very
similar to the $^{12}\rm{CO}/^{13}\rm{CO}$ ratio of 11.6 derived by
Wannier \&\ Sahai (1987) for CRL 618, which is the lowest they found
for any of the ten C and one S stars they investigated. The
$^{12}\rm{CO}/^{13}\rm{CO}$ ratios for the other stars in their sample
range from 14 to 48. In contrast, Kahane et al. (1992) derive
$^{12}\rm{CO}/^{13}\rm{CO}$= 18 and 3.2 for the 1-0 and 2-1 rotational
lines, respectively, for CRL 618. However, they adopt a {\it lower
limit} of 30 for $[^{12}{\rm C}/^{13}{\rm C}]$, which they derive from
the $[^{12}\rm{C}^{34}{\rm S}]/[^{13}{\rm C}^{32}{\rm S}]$ double
ratio. From carbon monoxide observations, Balser et al. (2002)
determined the $^{12}{\rm C}/^{13}{\rm C}$ ratio for 11 planetary and
protoplanetary nebulae and found values between 4 and 32, with a lower
limit of 4.6 for CRL 618 being at the lower end.

There are thus a number of uncertainties in 
the determination of the $[^{12}\rm{C}/^{13}\rm{C}]$ in CRL 618.
However, observations of C-bearing molecules other than 
carbon monoxide should be useful to get a clearer
picture since they might be subjected differently to
fractionation and photodissociation than the CO isotopomers,
where, according to Kahane et al. (1992) these processes
cancel each other.

  \acknowledgments
  The present study was supported by the Deutsche
  Forschungsgemeinschaft (DFG) via Grand SFB 494.  FW was partly
  supported by the National Science Foundation under Grant No.
  96-13716.

 \appendix

 \section{The HC$_3$N Partition Function}
   \label{sec-part}

   The partition function can be written as the product of its
   vibrational and rotational part:
   \begin{equation}
   \label{eq_qtot}
      Q=Q_v \cdot Q_r
  \end{equation} 

  In cold gas the contribution of vibrationally excited states can be
  neglected and for linear molecules the partition function is simply
  $Q=Q_r=1/3+kT/hB$. In hotter environments molecules are excited into
  vibrational states and $Q_v$ is given in LTE as
   \begin{equation}
   \label{eq_qv}
      Q_v = \sum_{v_1=0}^{\infty} \ldots \sum_{v_7=0}^{\infty}
            \exp \left( \frac{-hc\Sigma(v_i\omega_i)}{kT} \right)
            \prod_{i=5}^{7} (v_i+1)
  \end{equation}  
  where the sums can be evaluated analytically to yield:
   \begin{equation}
   \label{eq_qv-simple}
      Q_v = \prod_{i=1}^{7} 
      \left( 1-\exp \left( \frac{-hc\omega_i)}{kT} \right) \right)^{-d_i}
  \end{equation} 
  (Herzberg 1945). Here $v_i$ is the vibrational quantum number and
  $d_i$ denotes the degree of degeneracy of vibration $i$ with
  vibrational frequency of $\omega_i$.

\begin{figure}[p]
      \epsscale{1}
      \plotone{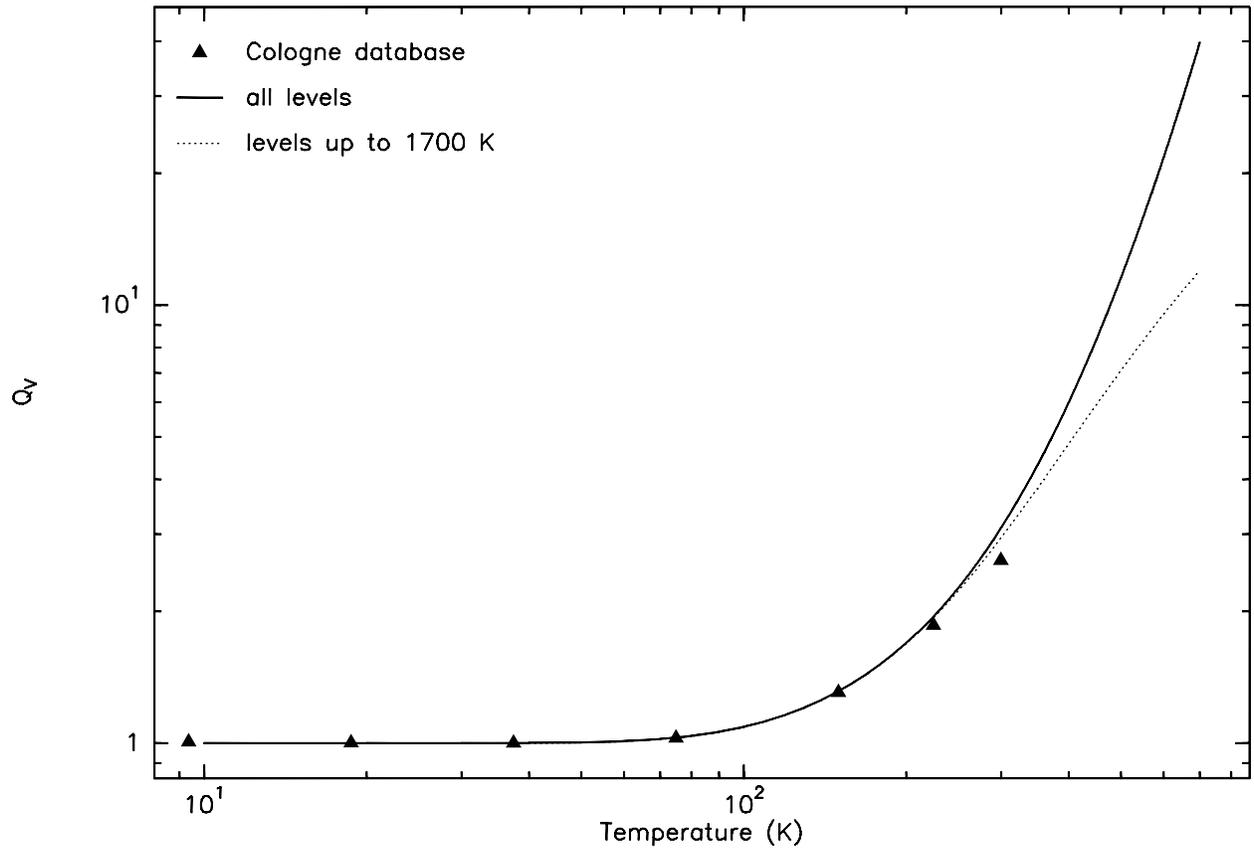}
    \caption{
      Vibrational part $Q_v$ of the partition function as discussed in
      Sect.~\ref{sec-part} compared with results from CDMS (M\"uller
      et al.\ 2001).
      }
   \label{hc3n-qv3}
  \end{figure}

  For non-LTE conditions, the higher levels might not be populated for
  two reasons: 1.) critical densities for the higher levels are
  generally larger so that the density might be too small to populate
  the level.  2.) Vibrationally excited states might be excited by
  infrared radiation fields. Higher levels correspond to shorter
  wavelengths in the NIR, where the radiation might be too weak to
  pump the levels. Hence, to estimate a lower limit for the partition
  function we computed Eq.~\ref{eq_qv} only up to the highest energy
  observed. The resulting lower and upper limits for the vibrational
  part of the partition function as a function of temperature are shown
  in Fig.~\ref{hc3n-qv3}, together with values from laboratory
  measurements published in the Cologne Database for Molecular
  Spectroscopy (CDMS, http://www.cdms.de/, M\"uller et al.\ 2001).

 \section{The Radiative Transfer}
   \label{sec-rt}

   To calculate the intensity $I_{\rm tot}$ at a certain position $R$
   and velocity $v$, we integrate over the line of sight $z$ at $R$:
   \begin{equation} 
     I_{\rm tot}(R)  = \int dz
                        \left[I(R,z) \, \exp(-d\tau(R,z))
                              +S(R,z) \, (1-\exp(-d\tau(z))) \right]
   \end{equation}
   with
   \begin{equation}  
      S(R,z) = \frac{h\nu}{\exp(h\nu/kT_{\rm ex})-1}.
   \end{equation}
   Here, $I(R,z)$ is the intensity integrated up to a position $z$ on
   the line of sight. $S$ is the source function at a position
   $(R,z)$. $T_{\rm ex}$ is either the kinetic temperature $T_{\rm
     K}$ of the molecular gas or the electron temperature $T_{e}$ of
   the embedded HII region. The optical depths at any position $(R,z)$
   is given as:
   \begin{equation} 
      d\tau(R,z) = \frac{c^2}{8\pi \nu^2} \, A \, n_u  \,
                   (1-\exp(h\nu/T_{\rm K}) \, \phi_0 
   \end{equation}
   with 
   \begin{equation}
      \phi_0 = \frac{2c\sqrt{\log(2)}}{\nu \sqrt{\pi}\Delta v} \,
               \exp\left( -4 \log(2) \, 
                   \frac{(v-v(R,z))^2}{\Delta v^2} \right)
   \end{equation}
    and the population of the upper states in LTE:
   \begin{equation} 
      n_u = \frac{g}{Q} \, n \, \exp\left(\frac{h\nu}{kT}\right).
   \end{equation}
   The optical depth through the HII region is:
   \begin{equation} 
      \tau_c = 0.0824 \, T_e^{-1.35} \, \nu^{-2.1} \, EM
   \end{equation}


\bibliography{}
\bibliographystyle{astron}


\newpage


\begin{table}
\begin{center}
  \caption[\CYAC\ $J=5-4$ Line Parameters]
    {    \CYAC\ $J=5-4$ line parameters. The brightness of emission lines is
         given in $T_{\rm MB}$, whereas absorption features are given
         in $T_{\rm MB}/T_{\rm C}$. Heavily blended lines are
         omitted.} 
  \label{param-100m}
  \vspace{.3cm}
\begin{tabular}{lcccr@{}lr@{}lr@{}l}
\hline \hline
Isotopomer & $(v_4,v_5,v_6,v_7)$ &$\ell$& $\nu$ & \multicolumn{2}{c}{$T_{\rm MB}$ or $T_{\rm MB}/T_{\rm C}$} & 
          \multicolumn{2}{c}{$v_{\rm LSR}$} & $\Delta v$ \\
         &     &  & (MHz) & \multicolumn{2}{c}{(K) or ()}  & \multicolumn{2}{c}{(\kms)}       &
                      \multicolumn{2}{c}{(\kms)}  \\
\hline
\CYAC\  &    $(0,0,0,0)$  &       & 45490.314e & $-$0 & .170\,(20) & $-$38 & .7\,(13)&  11&.4\,(24) \\ 
       &                 &       &            &    0 & .290\,(50) & $-$16 &. 6\,(6) &  14&.1\,(12) \\
       &    $(0,1,0,0)$  & $1e$  & 45494.714e & $-$0 & .120\,(20) & $-$27 & .2\,(17)&  14&.0\,(41) \\ 
       &    $(0,1,0,0)$  & $1f$  & 45520.454e & $-$0 & .064\,(14) & $-$28 & .1\,(7) &   6&.8\,(15) \\        
       &    $(0,0,1,0)$  & $1e$  & 45564.964e & $-$0 & .130\,(10) & $-$27 & .3\,(3) &   3&.5\,(9) \\ 
       &    $(0,0,1,0)$  & $1f$  & 45600.785e & $-$0 & .115\,(16) & $-$27 & .1\,(4) &   2&.7\,(5) \\  
       &    $(0,0,0,1)$  & $1e$  & 45602.171e & $-$0 & .180\,(16) & $-$31 & .8\,(4) &  12&.5\,(11) \\ 
       &    $(0,0,0,1)$  & $1f$  & 45667.550e & $-$0 & .240\,(40) & $-$30 & .8\,(4) &  10&.2\,(10) \\ 
       &    $(0,0,1,1)$  & $0$   & 45727.452e & $-$0 & .040\,(20) & $-$27 & .7\,(7) &   6&.6\,(11) \\ 
       &    $(0,0,1,1)$  & $2$   & 45729.306e & \\ 
       &    $(0,0,0,2)$  & $0,2$ & 45779.054e & $-$0 & .080\,(30) & $-$28 & .9\,(2) &   7&.3\,(3) \\ 
       &    $(0,0,0,3)$  & $1e$  & 45856.001e & $-$0 & .096\,(18) & $-$27 & .5\,(4) &   3&.7\,(7) \\ 
\hline \hline
\normalsize
\end{tabular}
\end{center}
\end{table}

\begin{table}
\begin{center}
  \caption[\CYAC\ $J=12-11$ Line Parameters]
    {\CYAC\ $J=12-11$
    line parameters. 
    $(0,0,0,0)_{wng}$ acounts for the line wings of the ground state
    transitions originating from the high velocity molecular outflow.}
  \label{param-30m-12}
  \vspace{.3cm}
\small
\begin{tabular}{llccr@{}lr@{}lr@{}l}
\hline \hline
Isotopomer & $(v_4,v_5,v_6,v_7)$ &$\ell$& Frequency & \multicolumn{2}{c}{$T_{\rm MB}$} or $T_{\rm MB}/T_{\rm C}$ & 
          \multicolumn{2}{c}{$v_{\rm LSR}$} & \multicolumn{2}{c}{$\Delta v$} \\
         &     &  & (MHz) & \multicolumn{2}{c}{(K) or ()}           & \multicolumn{2}{c}{(\kms)}       &
                      \multicolumn{2}{c}{(\kms)}  \\
\hline
\CYAC\         &  $(1,0,0,0)$      &       &109023.305  & $-$0&.220\,(30) & $-$27&.6(2) &   3&.7\,(3) \\ 
HC$^{13}$CCN  &  $(0,0,0,1)$      &  $1f$ &109125.744e & $-$0&.120\,(30) & $-$28&.5(4) &   3&.7\,(9) \\ 
HCC$^{13}$CN  &  $(0,0,0,1)$      &  $1f$ &109139.707e & $-$0&.150\,(30) & $-$29&.0(4) &   6&.2\,(8) \\
\CYAC\         &  $(0,0,0,0)$      &       &109173.634e & $ $0&.860\,(20) & $-$20&.4(1) &  17&.6\,(1) \\
              &  $(0,0,0,0)_{wng}$&       &109173.634e &    0&.100\,(10) & $-$14&.0(14)& 124&.0\,(50) \\    
              &  $(0,1,0,0)$      &  $1f$ &109244.222  & $-$0&.330\,(50) & $-$28&.9(2) &   5&.6\,(4) \\ 
              &                   &       &            & $ $0&.040\,(10) & $-$19&.6(8) &   7&.9\,(19) \\ 
              &  $(1,0,0,1)$      &  $1e$ &109306.704e & $-$0&.150\,(30) & $-$27&.1(4) &   2&.7\,(9) \\ 
              &  $(0,0,1,0)$      &  $1e$ &109352.781e & $-$0&.420\,(50) & $-$30&.4(3) &   6&.0\,(5) \\ 
              &                   &       &            & $ $0&.090\,(10) & $-$22&.2(6) &   9&.8\,(14) \\ 
              &  $(1,0,0,1)$      &  $1f$ &109469.409e & $-$0&.100\,(30) & $-$27&.5(4) &   4&.1\,(8) \\ 
\hline \hline
\normalsize
\end{tabular}
\end{center}
\normalsize
\end{table}

\begin{table}
\begin{center}
  \caption[\CYAC\ $J=15-14$ Line Parameters]
    {Same as Table~\ref{param-30m-12} for \CYAC\  $J=15-14$ }
  \label{param-30m-15}
  \vspace{.3cm}
\small
\begin{tabular}{llccr@{}lr@{}lr@{}l}
\hline \hline
Isotopomer & $(v_4,v_5,v_6,v_7)$ &$\ell$& Frequency & \multicolumn{2}{c}{$T_{\rm MB}$} or $T_{\rm MB}/T_{\rm C}$ & 
          \multicolumn{2}{c}{$v_{\rm LSR}$} & \multicolumn{2}{c}{$\Delta v$} \\
         &     &  & (MHz) & \multicolumn{2}{c}{(K) or ()}           & \multicolumn{2}{c}{(\kms)}       &
                      \multicolumn{2}{c}{(\kms)}  \\
\hline
HC$^{13}$CCN  &  $(0,0,0,1)$      &  $1f$ &136404.396e & $-$0&.166\,(21) & $-$28&.9\,(8) &   4&.9\,(12) \\ 
              &                   &       &            &    0&.060\,(10) & $-$19&.8\,(7) &   5&.6\,(12) \\
HCC$^{13}$CN  &  $(0,0,0,1)$      &  $1f$ &136421.848e & $-$0&.094\,(25) & $-$29&.7\,(22)&   5&.6\,(34) \\ 
              &                   &       &            &    0&.050\,(10) & $-$19&.7\,(10)&   7&.2\,(16) \\     
\CYAC\         &  $(0,0,0,0)$      &       &136464.411e &    0&.900\,(20) & $-$20&.6\,(1) &  16&.5\,(3) \\ 
              &  $(0,0,0,0)_{wng}$&       &136464.411e &    0&.200\,(20) & $-$21&.9\,(10)&  97&.6\,(33) \\ 
              &  $(0,1,0,0)$      &  $1f$ &136551.798  & $-$0&.306\,(39) & $-$29&.5\,(5) &   5&.3\,(9) \\ 
              &                   &       &            &    0&.100\,(10) & $-$20&.3\,(8) &   5&.9\,(16) \\ 
              &  $(0,0,1,0)$      &  $1e$ &136688.252e & $-$0&.336\,(45) & $-$31&.3\,(10)&   6&.2\,(16) \\ 
              &            &      &                   &    0&.140\,(20) & $-$19&.0\,(8) &   8&.8\,(13) \\ 
\hline \hline
\normalsize
\end{tabular}
\end{center}
\normalsize
\end{table}

\begin{table}
\begin{center}
  \caption[\CYAC\ $J=23-22$ Line Parameters]
    {Same as Table~\ref{param-30m-12} for \CYAC\ $J=23-22$ }
  \label{param-30m-23}
  \vspace{.3cm}
\small
\begin{tabular}{llccr@{}lr@{}lr@{}l}
\hline \hline
Isotopomer & $(v_4,v_5,v_6,v_7)$ &$\ell$& Frequency & \multicolumn{2}{c}{$T_{\rm MB}$} or $T_{\rm MB}/T_{\rm C}$ & 
          \multicolumn{2}{c}{$v_{\rm LSR}$} & \multicolumn{2}{c}{$\Delta v$} \\
         &     &  & (MHz) & \multicolumn{2}{c}{(K) or ()}           & \multicolumn{2}{c}{(\kms)}       &
                      \multicolumn{2}{c}{(\kms)}  \\
\hline
HC$^{13}$CCN  &  $(0,0,0,1)$      &  $1f$ &209137.396 &    0&.110\,(30) & $-$20&.6\,(8) &  10&.1\,(16) \\ 
HCC$^{13}$CN  &  $(0,0,0,1)$      &  $1f$ &209164.141 &    0&.130\,(30) & $-$21&.6\,(7) &   9&.1\,(12) \\ 
       \CYAC\  &  $(0,0,0,0)$      &       &209230.234 &    1&.120\,(50) & $-$20&.8\,(1) &  17&.6\,(4) \\ 
              &  $(0,0,0,0)_{wng}$&       &209230.234 &    0&.370\,(50) & $-$20&.6\,(10)&  94&.6\,(33) \\ 
              &  $(0,1,0,0)$      &  $1f$ &209362.113  &    0&.260\,(40) & $-$20&.3\,(4) &  11&.3\,(9) \\ 
              &  $(0,0,1,0)$      &  $1e$ &209573.178 &    0&.320\,(40) & $-$20&.9\,(4) &  10&.2\,(8) \\ 
HC$^{13}$CCN  &  $(0,0,0,2)$      &  $0$  &209597.019$^a$ &   0&.040\,(20) & $-$20&(2)&  9&(2) \\ 
              &  $(0,0,0,2)$      &  $2f$ &209628.074$^a$   \\ 
              &  $(0,0,0,2)$      &  $2e$ &209661.546$^a$   \\
HCC$^{13}$CN  &  $(0,0,0,2)$      &  $0$  &209634.224$^a$ & \\
              &  $(0,0,0,2)$      &  $2f$ &209663.206$^a$   \\
              &  $(0,0,0,2)$      &  $2e$ &209694.908$^a$   \\
\hline \hline
\end{tabular} 
\begin{flushleft}
$^a$ Line parameters were kept the same for all lines\\
\end{flushleft}

\end{center}
\normalsize
\end{table}


\begin{table}
\small
\centering
\caption[cont-fluxes]
        {Observed continuum flux densities of CRL~618.}
\vspace*{2mm}
\label{cont-fluxes}
\begin{tabular}{rrr}
\hline
Frequency & Flux & Error \\
    (GHz) & (mJy) &  (mJy)  \\
\hline
  4.5$^b$  &    26 &    3 \\   
 20.2$^b$  &   425 &    60 \\   
 24.7$^b$  &   490 &   90 \\   
 34.8$^b$  &   680 &  140 \\   
 34.8$^c$  &   740 &  140 \\   
 40.8$^b$  &   750 &   75 \\   
 45.6$^a$  &   800 &  200 \\ 
 47.2$^b$  &   840 &   85 \\ 
 82.3$^c$  &  1650 &  400 \\   
109.2$^a$  &  1800 &  200 \\ 
136.5$^a$  &  1650 &  300 \\ 
209.0$^a$  &  1350 &  340 \\ 
\hline 
\end{tabular}
\begin{flushleft}
$^a$ this study\\
$^b$ Thorwirth et al.\ (2002)\\
$^c$ Thorwirth (2001)\\
\end{flushleft}
\end{table}

\begin{table}
\small
\centering
\caption[fit-results]
        {Best-fit parameters for expanding envelope model. Note, that
         the variation in the \CYAC\ density parameters reflect the
         uncertainty in the partition function calculation.}
\vspace*{2mm}
\label{fit-results}
\begin{tabular}{lcr}
\hline
Description & Parameter & Value \\
\hline
Temperature  & $T_0$      & 560 K \\ 
             & $\alpha_T$ & --0.8  \\
Density      & $n_0$      & 500 -- 1000 \percc \\ 
             & $\alpha_n$ & --4 -- --5   \\
Velocity     & $v_{sys}$  & --24.2 \kms \\ 
             & $v_0$      & 3.3 \kms \\
             & $\alpha_v$ & 1.6   \\
             & $\Delta v$ & 4 \kms \\
$\rm ^{12}C/^{13}C$  & $X$ & 10 \\
\HII\ region & $T_e$      &  20900 K \\
             & $EM$       & $4.2\times 10^{10}$~cm$^{-6}$pc \\
             & $r_0$      & 0.11\arcsec \\
\hline 
\end{tabular}
\end{table}


\end{document}